# SPECTRe: Substructure Processing, Enumeration, and Comparison Tool Resource: An efficient tool to encode all substructures of molecules represented in SMILES

*Yasemin Yesiltepe [1,2], Ryan S. Renslow [1,2]\* and Thomas O. Metz [2]\**

1 The Gene and Linda Voiland School of Chemical Engineering and Bioengineering, Washington State University, Pullman, WA, USA.

2 Earth and Biological Science Division, Pacific Northwest National Laboratory, Richland, WA, USA

*Corresponding authors: Ryan Renslow (ryan.renslow@pnnl.gov); Thomas Metz (thomas.metz@pnnl.gov)

**Keywords:** substructure, fingerprint, SMILES, fragment, enumeration, chemical graphs, breadth-first search, depth-first search, fragment-based drug discovery, drug design

# ABSTRACT

Functional groups and moieties are chemical descriptors of biomolecules that can be used to interpret their properties and functions, leading to the understanding of chemical or biological mechanisms. These chemical building blocks, or sub-structures, enable the identification of common molecular subgroups, assessing the structural similarities and critical interactions among a set of biological molecules with known activities, and designing novel compounds with similar chemical properties. Here, we introduce a Python-based tool, SPECTRe (Substructure Processing, Enumeration, and Comparison Tool Resource), designed to provide all substructures in a given molecular structure, regardless of the molecule size, employing efficient enumeration and generation of substructures represented in a human-readable SMILES format through the use of classical graph traversal (breadth-first and depth-first search) algorithms. We demonstrate the application of SPECTRe for a set of 10,375 molecules in the molecular weight range 27 to 350 Da (≤26 non-hydrogen atoms), spanning a wide array of structure-based chemical functionalities and chemical classes. We found that the substructure count as a measure of molecular complexity depends strongly on the number of unique atom and bond types present, degree of branching, and presence of rings. The substructure counts are found to be similar for a set of molecules belonging to particular chemical classes and classified based on the characteristic features of certain topologies. We demonstrate that SPECTRe shows promise to be useful in many applications of cheminformatics such as virtual screening for drug discovery, property prediction, fingerprint-based molecular similarity searching, and data mining for identifying frequent substructures.

# INTRODUCTION

Molecular structures are formed of molecular blocks or substructures as chemical fingerprints of molecules, generally a moiety encapsulating chemical or biological functionality or representing the chemical class of a compound [1-3], and important molecular descriptor to interpret chemical properties of biomolecule [4]. Substructures are the fundamental determinant in our understanding of the mechanisms of the chemical or biological activities in human health related-, and environmental and agricultural samples [5-7]. The presence of certain building blocks in molecular structures can explain the formation of specific molecular interactions and/or provides identification of certain structural features [8-10]. Substructure-based researches are extensively used in chemical compound classifications [11], drug development processes [12], fragment-based drug discovery (FBDD) [13], *de novo* design of novel compounds [14], and identification of structural features [15].

In drug discovery processes, active compounds are identified by high-throughput screening, then they are modified and optimized to generate lead compounds [16], which might be expensive and requires time-consuming screening experiments [17]. Virtual screening (VS) provides rapid elimination of candidate molecules among millions of compounds found in chemically diverse VS libraries [18]. New drug-like compounds are generally generated combinatorially joining the sets of molecular building blocks and linkers, which are found in fragment libraries [19]. Therefore, it is crucial to have high-quality fragment libraries to generate novel chemical compounds for VS, as a diverse set of fragments leads to higher chances of driving novel hit candidate compounds [20]. Numerous fragment-based molecular synthesis tools can be used for computer-based *de novo* design of chemical compounds by assembling molecules from chemically feasible fragments such as CONFIRM [21], AutoGrow [22], and eSynth [23].

The first comprehensive reviews of fragment-based drug discovery (FBDD) were published in 2004 [24, 25], and ever since, FBDD has become a significant paradigm in designing of fragment libraries, discovering of novel drug candidates, analyzing fragment sets [26], and enabling efficient exploration of chemical space [27]. It is an alternative starting point for the discovery of high-quality lead candidates, and a useful approach that to track the biologically significant interactions at the molecular state. It is challenging to apply the FBDD approach to complex biological targets [30] because it is based on identifying only small chemical fragments (non-hydrogen atoms<20, molecular weight<300 Da) [28]. Those fragment counts are considerably lower than the number of drug-like molecules, which are typically discovered by another drug discovery approach, high-throughput screening [29]. Higher molecular complexity dramatically increases the probability of observing more useful interactions [31], and high-throughput screening is a commonly used approach to monitor more complex and larger biological interactions.

A number of fragmentation tools are available to efficiently break down molecules to nonredundant fragments such as molBLOCKS [32], Fragmenter [33], *e*MolFrag [34] applying retrosynthetic combinatorial analysis procedures (Recap rules - 11 chemical bond types derived from common chemical reactions [35] to cleave a molecule) and breaking retro-synthetically interesting chemical substructures (BRICKS, RECAP bond type criteria expanded to 16, derived from the chemical environment of each bond type and the surrounding substructures [36]).

As a cost-effective complement to high-throughput screening, VS provides efficient exploitation of chemical space for drug discovery organizing molecules using numerical descriptors in a form of a vector or fingerprint [37]. Molecules are represented as bitvectors representing chemical fingerprints. In these bitvectors, each bit represents the presence or absence of a substructure, specific structural feature and molecular property. There has been a comprehensive work on developing efficient algorithms and strategies in many uses/areas of cheminformatics such as property prediction [38], fingerprint-based molecular similarity searching [39], data mining in finding of frequent substructure [40], classification of organic molecules by molecular quantum numbers (counting atoms and bonds, polarity and topology) [41].

In rational drug discovery and chemical genomics, fingerprint-based molecular similarity searching approach has been applied frequently by computational and medicinal chemists. Such efficient database searching methods provide identifying common structural patterns shared by sets of bioactive compounds showing some similarity to each other [42]. Another application of similarity assessment is the molecule diversity analysis in industrial pharmaceutical research to generate structurally representative clusters for starting new biological assays and/or to increase the diversity of a corporate database by "rational" acquisition of compounds [43]. The similarity of molecules is quantified by many common similarity metrics depending on the fingerprint method such as Euclidean distance [44] and industrial standard the Tanimoto coefficient. However, Tanimoto coefficient is in the range of 0 to 1 regardless of the length of the fingerprint, therefore losing the ability of being a distinctive measure as fingerprints get longer [45].

Among the structural descriptor-based strategies, structural keys-based fingerprints mostly use 2D molecular graphs capable of storing atoms and bonds information. Each provides a different aspect of the molecule depending on the presence of the pre-defined fingerprints/fragments in the compound. These pre-defined fingerprints are from the supplied list of structural keys. Most commonly used fingerprints are PubChem fingerprints [48] with 881 structural keys used by PubChem [49] for similarity searching and neighboring, and implemented in ChemFP [50] and CDK [51, 52], MACCS [53] with 166 and 960 structural keys available in Open Babel [54], BCI fingerprints [55] with 1052 structural keys available in BCI toolkits, and TGD pharmacophoric fingerprints [56] available in MOE software package [57]. These fingerprints cover most of the chemical features for drug discovery and pharmaceutical science, however they are strictly based on the presence or absence of the user-defined structural keys in compounds and only useful when used with those that are already likely composed of similar fragments.

To address the lack of generality of the structural keys, topological fingerprints analyze all the fragments of compounds based on following paths and create the fingerprints by hashing every one of these paths into a fixed-length bitvector. Path-based fingerprints are extensively used for fast substructure searching and filtering. The Daylight fingerprint [58] encodes paths up to 2048 bits. Each substructure is mapped to a bit position, instead of being assigned to specific bits. It removes the limitation of predefined substructures and provides enumeration of all fragments of a molecule. However, these fingerprints are prone to has collisions, resulting in representation of more than one feature with the same bitvector [41]. Regardless of the molecule size, paths are encoded to a maximum of 2048 bits. Moreover, fingerprints are implemented following linear paths, not covering the rings, bridges, and branches, and up to only an adjusted length, i.e. certain number of bonds. FP2 is a path-based fingerprint, available in Open Babel, an open-source chemical informatics toolbox [54]. FP2 consists of 1024 bits and indexes linear segments of the maximum length of 7 atoms. These bit-vectors structural fingerprints are not human readable, but optimal to encode chemical and topological characteristics of small molecules (linear molecules having 1-7 non-hydrogen atoms) and widely used for the small molecule comparisons and for fragment searches using structural features. However, they do not provide fragments having larger than 7 non-hydrogen atoms, which might lead missing of most of the critical chemical features for drug discovery.

To this end, and to help researchers in pharmaceutical science bring text-based and human readable fragments in canonical SMILES form, here we introduce a new open-source python-based tool, the SPECTRe (Substructure Processing, Enumeration, and Comparison Tool Resource) designed for efficient enumeration of substructures of a given molecular structure employing graph-based search algorithms. In this paper, we describe the implementation of graph-based theory using efficient search algorithms, provide a supporting tutorial, demonstrate its use for a large set of small organic molecules relevant to the mainstream FBDD community. The functionality of SPECTRe is shown for a set of metabolites in this paper, but it can be applied to several different families of molecules.

**MATERIALS AND METHODS**

**SPECTRe – Python Package**

SPECTRe, shown schematically in **Figure 1**, is an open source python package which provides straightforward fragmentation, decomposition, and enumeration of substructures of a given molecular structure. It is a user-friendly tool that only requires a list of molecules in one of the following formats: A) Simplified Molecular Input Line Entry System (SMILES) [59, 60] B) International Union of Pure and Applied Chemistry (IUPAC) International Chemical Identifier (InChI) [61], or C) MDL MOLfiles [62]. If the user chooses the options A or B, then SPECTRe calls Open Babel to convert InChI or SMILES to InChIKey. If the user choses option C, then SPECTRe uses the associated MOL files for naming files. Once the user prepares the input file containing the list of molecules, then SPECTRe starts to generate the substructures and writes them to two output files, one in SMILES form and the other in MDL SDF file format. The wall-clock time, including input/output (I/O) operations and processing/operating the SPECTRe codes, will depend on the local computer performance.

SPECTRe can be installed locally and is available at github.com/pnnl/spectre.

A supporting tutorial is provided to guide users in preparation of input files and describing the steps to run SPECTRe on two small molecules, methanol ($CH_3OH$; SMILES: CO), and toluene ($C_7H_8$; SMILES: Cc1ccccc1). These molecules are chosen because they are small in size (i.e. number of non-hydrogen atoms), different in shape, and do not require much computational power and time (less than 1 minute expected). The tutorial includes step-by-step instructions for preparing input files and running SPECTRe for the first time and details for the installation of Open Babel. SPECTRe does not require any additional python modules and packages.

**Algorithm**

A molecule is described by a 2D chemical graph consisting of a set of nodes (atoms) and a set of edges (bonds). A set of features for each atom is attached to the corresponding vertex in the molecular graph, such as atom type (C, O, N, S, …), state of charge (-1, -2, 0, … ), mass difference (-1, 0, 1, 2, …), hydrogen count (1, 2, … i.e. explicit hydrogens), stereo configuration, and valence. Similarly, a set of features for each bond is attached to the graph such as bond type (1=single, 2=double, …), vertex indices sharing that bond, and bond topology (1=ring, 2=chain, …). An adjacency list (connection table) is built to represent this graph. Only non-hydrogen atoms are considered (i.e. implicit/terminal hydrogens are omitted unless shown explicitly). The adjacency list is constructed from MOLfiles that contain information about the atoms, bonds, and their connectivity.

Substructures are identified in two stages. In the first stage, only the paths consisting of a straight chain of node are found, where each node is connected to every other in the molecule (i.e. a sequence of linearly connected atoms). The Breadth-first search (BFS) [63, 64] technique is applied to find every possible path that exists in the molecule. BSF is an extensively used graph traversal algorithm to identify the paths from the starting node up to the destination node. It starts from an arbitrary node and visits all of its first-degree neighboring nodes. Then, it moves to its one level deeper neighbors repeating the same procedure until no connecting node is left to visit.

In the second stage, branched paths are searched using the Depth-first search (DFS) algorithm [65, 66]. This time, an opposite strategy is used, which explores a node branch from the starting node as far as possible until there is no longer a connecting node left. As a result of this stage, the branches are mapped to the chain paths at possible positions. Any duplicate path is removed at the end of both stages; thus, no substructure is listed twice within a single run. Obtained paths are converted to canonical SMILES where each structure is always represented by the same SMILES string using Open Babel [54]. Different paths may still result in the same canonical SMILES strings, and SPECTRe cleans these duplicates in a final step.

**Demonstration Set**

We demonstrate the utility of SPECTRe for a set of 10,375 molecules taken from HMDB and in the molecular weight range of 27 to 350 Da (<=26 non-hydrogen atoms). Our set spans a wide array of structure-based chemical functionalities and chemical classes consisting of inorganic (51) and organic compounds including organophosphorus (2), organometallic (12), organosulfur (201), organoheterocyclic (2016) compounds, organohalogens (42), benzenoids (1675), phenylpropanoids (2160), lipids (1541), hydrocarbons (127), alkaloids (107), lignans (11), acetylides (1), nucleosides (10), acids (1280), oxygen (993), and nitrogen compounds (151) with the number of compounds written in parentheses. The classes are determined by the hierarchical chemical classification scheme, ClassyFire [67]. The complete set of molecules and their substructures are given in the Supporting Information (SI).

**Computational details**

All computational processes were performed using the Cascade high-performance computer (1440 compute nodes, 23,040 Intel Xeon E5-2670 processor cores, 195,840 Intel Xeon Phi 5110P coprocessor cores, and 128 GB memory per compute node [68]), in the Environmental Molecular Sciences Laboratory (a U.S. national scientific user facility) located at Pacific Northwest National Laboratory.

**RESULTS AND DISCUSSION**

**Substructure count as a measure of molecular complexity**

The lack of a common definition of the term 'molecular complexity' has sourced from various perceptions of complexity as assessed by both intrinsic chemical skeletal properties (i.e. molecular topology generally based on graph theory and mostly measured with elemental diversity and the number of connections in a chemical graph) and extrinsic molecular properties such as the ease of molecular synthesis and accessibility [69, 70]. As a measure of complexity, topological indices have become popular and resulted in numerous applications in pharmaceutical developments and drug design [71, 72]. In addition, many crowd-sourced mathematical approaches have been introduced for quantifying molecular complexity, such as total number of connected sub-graphs [73-75], total walk count [76], assessment of symmetry [77], counting of skeletal features (e.g. number of rings, chiral centers, single/double/triple bonds, heteroatoms [78], functional groups, and electronegativity etc. [79]), and summation of the complexity of every atom environment in a molecule [80]. In our analysis, we systematically investigated the effect of major contributors (e.g.

branching, cycled paths, bond type, atom type, and molecular class) to substructure count for the demonstration set (10,375 molecules).

**Figure 2** shows the number of unique substructures of each molecule (in logarithmic scale) as a function of molecule size (molecular weight) for the demonstration set. As with many other complexity measures, substructure count increases with the molecule size, i.e. number of atoms and bonds, and the degree of branching. For example, we investigated the effect of branching on the substructure counts for a set of 63 alkanes ranging in size up to 25 carbon atoms through controlled experiment/test using a set of linear compounds and compounds having up to 4 branches. Our criteria were to have molecules consisting of only carbons and no rings for comparisons because the presence of different atom types and any cycled paths increases the chances of having more substructures. The least count of substructures was found for the linear molecules, and the substructure counts increased exponentially with the degree of branching. Note that 93% of the demonstration set has at least one branch, so branching is one of the key factors affecting the substructure count in the majority of the dataset. See SI for details.

Besides branching, another key parameter significantly affecting the substructure count is the presence of cycled paths, i.e. rings and bridges over rings, especially considering that 80% of the demonstration set have at least 1 cycled fragment. It is observed that the substructure counts are higher for cycled molecules than linear molecules using a set of cycloalkanes (6 compounds) and linear alkanes (21 compounds) which consist of only carbons and single bonds and no branches. See SI for details.

Along with branching and cycling, the substructure counts strongly depended on the presence of different types of bonds (i.e. single, double, triple …), as demonstrated for two sets of linear hydrocarbons saturated (72 compounds consisting of only single bonds) and unsaturated (47 compounds with some number of double and/or triple bonds) compounds. It was clearly observed that the substructure counts of unsaturated hydrocarbons are higher than those of saturated ones. See SI for details.

Another factor increasing the substructure counts is the presence of multiple types of atoms (i.e. Br, Cl, N, O, S, …) in the molecules. We compared two sets of linear molecules consisting of only single bonds; one with one type of atom (i.e. hydrocarbons), and the other with multiple types of atoms (i.e. alcohols). The latter group has more unique substructures than the former does. See SI for details.

Furthermore, the substructure count of each molecule changes with the class of the molecule (alkyl groups, primary/secondary/tertiary alcohols, aldehydes, and ketones etc. as defined by ClassyFire), depending on the characteristic of certain topologies. For example, lipids have more substructures than phenylpropanoids, and alcohols have more substructures than nitrogen compounds. The distribution of substructure counts was found to be similar within a single class of molecules. Villar et al. provided the analytic expressions for the numbers of substructures as a function of fragment and molecule size for many molecule classes [82]. See SI for details.

**Computational complexity**

Enumeration of sub-graphs is essentially an algorithmic problem using $O(n^3)$ elementary arithmetic operations for compounds comprising up to n non-hydrogen atoms [81]. We found that it is feasible to find substructures for molecules with less than 21 non-hydrogen atoms via SPECTRe. As the molecular graphs get more branched and increase in size, the complete and systematic enumeration of all substructures becomes challenging. A substantial number of chemical compounds are unfortunately formed of branched sub-graphs and cycled paths. For example, 94% and 97% of the demonstration set have more than one type of bond and one type of atom, respectively. The ratios of molecules of the demonstration set with rings and branches are 97%, and 90% of those molecules consist of multiple types of bonds and atoms. In other words, only 0.27% of the demonstration set are linear molecular structures with one type of bond and atom.

Substructure searching, molecular enumeration and molecular similarity searching is a non-polynomial-complete (NP-complete) problem, which means that it gets computationally expensive at a rate higher than exponential as the number of atoms and/or bonds considered get higher [46, 47]. In our analyses, the computational time for finding linear paths (using BFS) and branched paths (using DFS) similarly both increased faster than exponential with respect to non-hydrogen atom counts of the molecules. It takes 70, 1000, and 400,000 times more time finding branches than finding linear paths for molecules with less than or equal to 10, 20, and 27 non-hydrogen atoms, respectively. This sharp increase depends on a number of factors such as bridges over rings, multiple connected rings, overlapping of rings, shared walls between rings, and branching.

The side processes (i.e. attaching the generated paths in MOL formats (SDF), conversion of MOL to canonical SMILES followed by the removal of duplicated structures) are performed using the Linux utilities and Open Babel, as explained in Materials and Methods. Similarly, the computational times for these processes scale exponentially, depending on the molecule size and the factors (e.g., branching, cycled paths, bond type, atom type, and molecular class).

The computational costs of finding substructures for the demonstration set are given in the SI.

**Enrichment of largest substructures**

The maximum common substructure (MCS), the largest substructure between two compounds [42, 83] and/or multiple molecules [84] [85], is a commonly used metric for chemical similarity searching and grouping compounds sharing the same pattern. Different from fingerprints or structural keys, MCS offers larger patterns that appear in a set of compounds, not only the special cases defined in fragment libraries. However, it requires the exact match, which is a graph-based similarity problem and can be computationally very demanding when the molecule set consists of closely related complex structures. It is essential to find an efficient algorithm to minimize the computer time of finding MCS, which is an extremely complex search process. However, assessment of the best MCS algorithms to minimize the computational demand is beyond the scope of this study.

In drug discovery and chemical genomics, it is critical to find common structural patterns shared by sets of small molecules for measuring structural similarities and clustering. MCS is also commonly used for searching the largest substructures with unknown property value shared by a set of similar molecules for the assessment of the toxicity of the target chemical [86]. Recently, we applied SPECTRe [87] to a set of blinded complex mixtures from a study focused on testing current metabolite identification techniques for the U.S. Environmental Protection Agency's (EPA) Non-Targeted Analysis Collaborative Trial (ENTACT) [88]. We found the substructures of 1188 compounds in the molecule size range of 4 to 38 non-hydrogen atoms with the molecular weight range of 75 to 776 Da. The focus of our analysis of these compounds using SPECTRe was to investigate their trends in observability during mass spectrometry-based analysis and based on various properties including molecular substructures. We identified the six largest common substructures representing 55% of the molecules using the following criterion: The six fragments were not the sub-fragments of each other and not shared by the same molecule. Thus, we clustered the molecules into six groups where each molecule can belong to only one group and cannot have the fragment of another group. The compounds were also grouped using the predicted chemical properties, artificial intelligence (AI)-based chemical space, and chemical class ontology, and then were compared and investigated for trends in observability. We demonstrated an example use of SPECTRe through such a substructure analysis to cluster compounds based on the criterion of searching the MCS(s) found in the compounds. We backtracked to the source compounds of MCS(s) to identify the common and unique features in their structures. Thus, SPECTRe has the potential to be used in studies finding the critical features of molecules related to observability or other relevant properties such as toxicity and ligand affinity.

**Substructure analysis in *de novo* molecular design**

One application of substructure analysis is in *de novo* molecular design. Combining substructure assessment with quantitative structure activity relationship (QSAR) models can locate particular substructures that are critical for the desired interaction at a binding site of interest.[89, 90] Furthermore, substructure analysis is useful in identifying common molecular moieties among molecules known to interact with particular receptors and in assessing structural similarity between molecules (e.g., similarity assessment between experimentally confirmed molecules and de novo generated molecules). Such information is used during the drug discovery process to identify molecular scaffolds that represent novel families of lead compounds.

Recently, we applied substructure analysis to the deep learning-based generation of molecules targeting the phencyclidine (PCP) site of the N-methyl D-aspartate receptor (NMDAR) to identify patterns in known active compounds and assess the similarity of computer-generated molecules [91]. First, we curated a dataset of known PCP site antagonists and identified the most common unique substructures found in these compounds. The dataset has 728 compounds with 8 to 35 non-hydrogen atoms with the range of 112 to 579 Da. This dataset was used to seed our deep learning software DarkChem [92] to generate potentially novel compounds with, ideally, high diversity compared to our starting set. We found that among the twenty most common substructures found in at least half of the antagonist dataset, each substructure was contained in no more than 0.25% of the computer-generated molecules. We then filtered the generated compounds using a combination of ligand- and structure-based methods until we obtained a finalist set of twelve. Each finalist contained only between one and eight of the twenty most common substructures. We concluded that we had indeed achieved a high degree of diversity between our seed and generated molecules.

**R group decomposition and computer-based *de novo* design of chemical compounds**

A detailed substructure analysis can lead to understanding the structure–biological activity relationship in synthetic or naturally occurring drugs and how any changes in the scaffold structure can alter the biological and pharmacological activities. We applied SPECTRe to a library of 388 known fentanyl analogues curated from eight different sources [67, 93-99]. SPECTRe decomposed the fentanyl molecules around the predefined scaffold member, piperidine, and extracted and labeled the substituents using the predefined connecting linkers. The library was then expanded with an additional 61,063 compounds by executing a substructure search of the piperidine subunit of the fentanyl molecule [97] through a database of 2.5 billion small molecules (combined of 40 publicly available databases). We searched through the newly expanded library for the most common substructures of fentanyl analogues. From these substructures, we hope to be able to more quickly and accurately identify new potential fentanyl analogues from unknown substances.

In the second stage of the application of SPECTRe, we performed a structure-based molecule design to enhance the chemical diversity of large collections of fentanyl-like compounds. We generated fentanyl molecules by combinatorically joining the sets of molecular building blocks (i.e. the substituents found in the first stage) to the fixed scaffold at the connecting linkers. At the end, we generated 1.7 trillion new fentanyl molecules. We hope this novel application will enrich the libraries of computer-based generated chemical compounds.

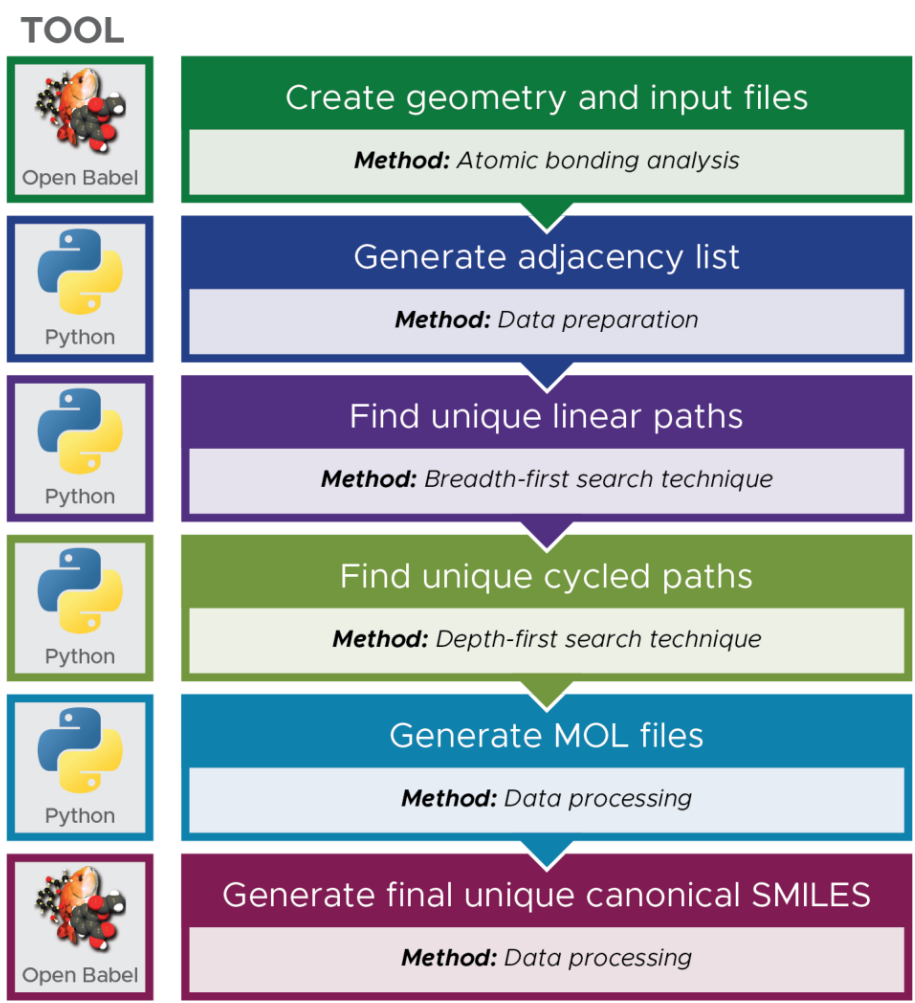

**Figure 1 The workflow of SPECTRe.** Please see github.com/pnnl/spectre

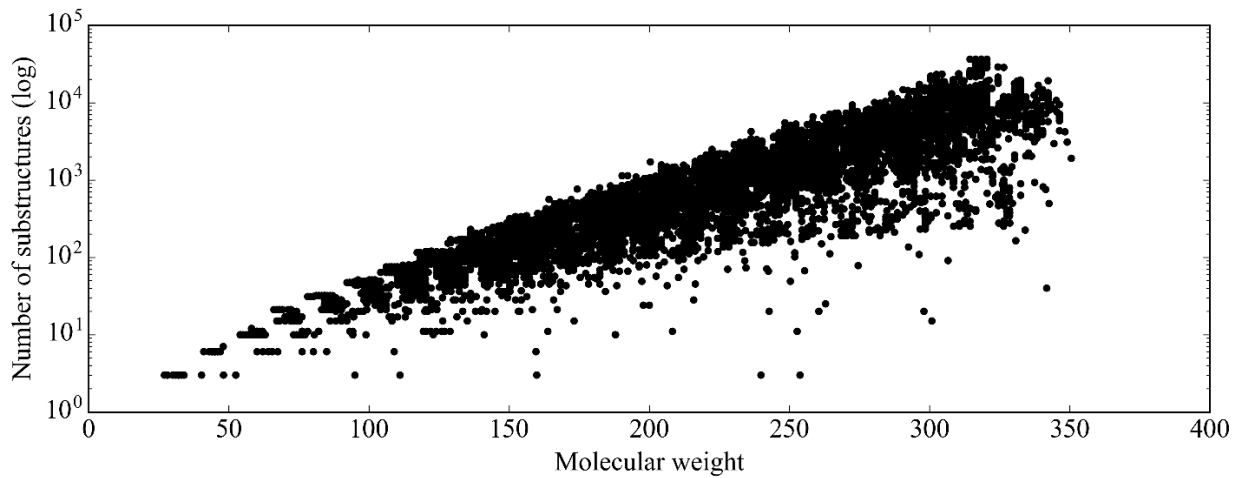

**Figure 2 The distribution of substructure counts of the molecules in the demonstration set (10375 molecules) vs. molecular weight (0-350 Da)**

**CONCLUSIONS**

SPECTRe is a user-friendly tool to provide all substructures in a given molecule structure with any molecule size. SPECTRe employs classical graph traversal algorithms for efficient enumeration and generation of substructures. SPECTRe writes the data to appended MDL MOL files, and then converts them to SMILES format which is the most commonly used and compact representation of molecules.

We show that SPECTRe is a promising tool for molecule fragmentation, decomposition, and substructure enumeration and could be used in many studies related to *de novo* molecular design, fragment-based drug discovery, classification of datasets, and drug development processes. We hope SPECTRe will contribute to expansion of the fragment libraries and their use in generation of novel compounds. It has the potential to aid in analyzing substructure sets, discovering novel drug candidates, and comprehensive and high-throughput exploration of chemical space.

SPECTRe can be run seamlessly on any local computer.

**FIGURES**

Figure 1 – General workflow of SPECTRe

Figure 2 – The substructure counts of the molecules in the demonstration set vs. molecular weight

AUTHOR INFORMATION

**Corresponding Author**

Ryan S. Renslow (ryan.renslow@pnnl.gov)

Thomas O. Metz (thomas.metz@pnnl.gov)

**Author Contributions**

The manuscript was written through contributions of all authors. All authors have given approval to the final version of the manuscript.

**Funding Sources**

This work was supported by the National Institutes of Health, National Institute of Environmental Health Sciences grant U2CES030170. Pacific Northwest National Laboratory (PNNL) is operated for the U.S. Department of Energy by Battelle Memorial Institute under contract DE-AC05-76RL01830.

**Notes**

Any additional relevant notes should be placed here.

ACKNOWLEDGMENT

Conflict of Interest: none declared.

ABBREVIATIONS

FBDD, fragment-based drug discovery; NP-complete, non-polynomial-complete; VS, virtual screening; BFS, Breadth-first search; DFS, Depth-first search.